%Paper: astro-ph/9309047
%From: Yoshiaki Sofue <sofue@sof.mtk.ioa.s.u-tokyo.ac.jp>
%Date: Wed, 29 Sep 93 17:25:52 JST

%%%%%%%%%%%%%%%%
%% Definitions:
%%%%%%%%%%%%%%%%
\def\sect{\vskip 2mm \centerline}
\def\r{\hangindent=1pc  \noindent}
\def\re{\hangindent=1pc  \noindent}
\def\ref{\hangindent=1pc  \noindent}
\def\cen{\centerline}

\def\v{\vskip 1mm}

\def\endpage{\vfil\break}

\def\noi{\noindent}
\def\kms{km s$^{-1}$}

\def\deg{$^\circ$}

\def\tmb{$T_{\rm mb}$}

\def\ico{$I_{\rm CO}$}

\def\Msun{M_{\odot \hskip-5.2pt \bullet}}
\def\msun{$M_{\odot \hskip-5.2pt \bullet}$}

\def\halpha{H$\alpha$}

\def\Deg{^\circ}
\def\deg{$^\circ$}

\def\co{$^{12}$CO($J=1-0$)}

\def\htwo{H$_2$}

\def\pa{PASJ}

\def\apj{ApJ}
\def\apjl{ApJ Letters}

\def\aj{AJ}
\def\aa{A\&A}

\def\aas{A\&AS}

\def\araa{ARAA}

\def\kluwer{Kluwer Academic Publishers, Dordrecht}
\def\pasp{PASP}

\def\so{Sofue, Y.}

\def\wa{Wakamatsu, K.}

\def\iaut{{\it Institute of Astronomy, University of Tokyo, Mitaka, Tokyo 181}}

\def\gifu{{\it Department of Physics, Gifu University, Gifu 501-11}}
\def\kiso{{\it Kiso Observatory, University of Tokyo, Kiso-gun, Nagano 397-01}}
%submitted to PASJ 1993.5.28
%revised 1993.9.28
\cen{\bf Face-on Barred Spiral Structure of Molecular Clouds in M31's Bulge}
\v\v
\cen{Y. Sofue$^{1,2}$, S. Yoshida$^1$, T. Aoki$^1$, T. Soyano$^1$, M.
Hamabe$^1$, and K. Wakamatsu$^3$}
\cen{1. \kiso}
\cen{2. \iaut}
\cen{3. \gifu}

\v\v
\cen{\bf Abstract}\v

By image processing and color excess analyses of $B, ~V, ~R$, and $I$-band CCD
images of the central region of M31 taken with the Kiso 105-cm Schmidt
telescope, we found a bar of 200 pc length.
We also found a more extended ``face-on" spiral feature of dark clouds, which
appear to be connected to this bar and is probably an out-of-plane structure.
The \co-line emission was detected from the dark clouds using the Nobeyama 45m
telescope, and the virial mass of individual clouds was estimated to be  $\sim
10^6\Msun$.
The CO intensity-to-virial mass ratio is anomalously small, and so is the color
excess-to-mass ratio, which indicates a low temperature and a small amount dust
per unit gas mass in the cloud, respectively.
This implies that the molecular clouds in the central few hundred pc of M31 are
more ``primeval'', suggesting a low star-formation and low dust supply.
We propose a possible scenario for the origin of the face-on spiral containing
such gas on the basis of a ram-pressure accretion model of stripped gas clouds
from the companion galaxies.

\v

\v\v\sect{\bf 1. Introduction}\v

Optical photometry of the central region of M31 showed an ellipsoidal bulge
without indication of disk structure (Kinman 1965).
Extensive surveys for optical dark clouds in the central region of M31 have
revealed neither a disk structure nor spiral arms that are related to the major
disk of M31 (Johnson and Hanna 1972; Hodge 1980; Gallagher and Hunter 1981;
McEroy 1983).
On the other hand, CCD imaging of the \halpha emission of the M31's bulge
revealed peculiar spiral patterns of ionized gas, which suggested an off-plane
ejection of gas driven by wind from bulge stars (Ciardullo et al. 1988).
Far-infrared observations with the IRAS indicated a central concentration of
warm dust (Habing et al. 1984).

In spite of the indication of an interstellar activity,  no significant
detection of  molecular and atomic hydrogen gases has been reported, indicating
a devoid of interstellar hydrogen in the central region, from CO-line
observations (Combes et al. 1977; Stark 1985; Koper et al. 1991) and  HI
observations (Roberts and Whitehurst 1975; Brinks and Bajaja 1986).
In this respect, the central region of M31 exhibits an early-type
characteristics (Gallager and Hunter 1981).
These facts  put M31 among exceptions in which little molecular gas is present
in the nuclear region, not alike other Sb galaxies like the Milky Way (Dame et
al 1990) or NGC 891 (Garcia-Brillo 1992; Sofue and Nakai 1993).

In this paper, we analyze CCD images in the $B, ~V, ~R$, and $I$-bands of the
M31's bulge by applying contrast-enhancement techniques, in order to study
detailed morphology in the central region.
We also report the detection of CO line emission toward several dark clouds,
and discuss their mass and kinematics.

\sect{\bf 2. Optical Bar and Dark Clouds}

\v\noi {\it 2.1.  CCD Images}

CCD images filtered in the $B, ~V, ~R$, and $I$-bands were taken on December
26, 1992  with the 1.05/1.50 m Schmidt telescope of the Kiso Observatory at the
f/3.0 primary focus.
The detector was the TC215-1K (=KCCD2; Kiso CCD No.2), which is a $1000 \times
1018$ device.
The pixel size was $0''.752$, which corresponded to 2.5 pc at 690 kpc distance
of the M31 center, so that the frame covered the central $12'.53\times 12'.76$
($2.52\times 2.56$ kpc$^2$).
The seeing size was estimated to be about $4''$ (=13 pc).
We obtained several frames with different exposure times at each band, and
flat-fielding was applied to each frame.
The data were added so that images we use have total exposure times of 570,
110, 55 and 40 seconds for the $B, ~V, ~R$, and $I$-bands, respectively.
Intensity calibration was made by using star images of the SA95 standard field.
Reduction was made by using the standard IRAF package.

The sky subtraction was rather ambiguous because of the extended emission of
the M31 disk.
However, usual contrast-enhanced (background-filtered) images, which we use to
identify dark clouds, are not affected by the ambiguity of sky subtraction,
since the background extended emission is subtracted.
In order to estimate color indexes, sky subtraction was necessary, and we
simply subtracted the darkest sky at the SE and NW corners 9$'$ from the
nucleus, which yielded intensity excess over that of the darkest sky in the
same frames.
The central region of the bulge within $2'$ of the nucleus, in which we are
interested, was found to have CCD counts an order of magnitude greater than
that at this darkest corner, and this simple subtraction resulted in an error
of about 5 \% of intensity in each band in the sense of underestimate of
intensity.
We actually compared the results (e.g., color excesses) with those obtained
from images without applying the sky subtraction, but found almost no
significant difference.
We apply contrast enhancing techniques to produce differential intensity
distribution, in which the error is reduced to about 2\% or 0.02 mag..

The color excess of a cloud is defined as
$$
E(B-V)=(B-V)_{\rm cloud}-(B-V)_{\rm bg}
=(B_{\rm cloud}-B_{\rm bg})-(V_{\rm cloud}-V_{\rm bg}),
$$
where the subscripts `cloud' and `bg' denote values for the cloud and its
surrounding smooth background, respectively.
If there exist errors of $\delta_{\rm B}$ and $\delta_{\rm V}$ in intensities
($I_{\rm B},~I_{\rm V}$) (such as due to ambiguous sky subtraction), the error
involved in color excess is estimated to be
$$
\delta_{\rm E}
\simeq 2.5
\lbrack
 (\delta_ B/I_{B,{\rm cloud}} - \delta_B/I_{B, {\rm bg}})^2
+(\delta_V/I_{V, {\rm cloud}} - \delta_V/I_{V, {\rm bg}})^2
\rbrack ^ {1/2}
$$
$$
\sim 2.5 \lbrack (\delta\Delta/I_{\rm bg}^2)_B^2 + (\delta\Delta/I_{\rm
bg}^2)_V^2 \rbrack ^{1/2},
$$
where  $\Delta = I_{\rm cloud}-I_{\rm bg} \ll I_{\rm bg}$.
Hence, the error in $E$ is estimated to be
$ \delta_{\rm E} \sim 0.009$  mag..
Errors due to read-out noise of the chip are estimated to be much smaller.

\v\noi{\it 2.2. An Optical Bar}

The original images do not clearly show up fine features, which are embedded in
the bright central bulge.
Hence,  we applied contrast-enhancing techniques as described in Sofue (1993)
in order to subtract the background diffuse emission.
First, we made use of the background-filtering technique (BGF; Sofue and Reich
1979), which is an equivalent method to the unsharp-masking method:
We subtracted bright stars and the nucleus, and convolved the resultant image
with a Gaussian function with HPBW=10 pixels.
Then, the original image was divided (not subtracted) by the smoothed image,
yielding a BGF image.
This is similar to the unsharp-masking method, but should rather be called an
``unsharp-dividing" method.
The unsharp-dividing gives contrasted-enhanced ratio of intensity measured in
magnitudes, and is convenient to discuss such features like extinction by dark
clouds silhouetted against background, and  the result does not depend on the
background intensity.
(Note, on the other hand, that unsharp-masking gives subtraction residuals,
which depends on the subtracted background intensities, so that a cloud would
appear darker if it is in front of bright background, and vise versa.)
Fig. 1 shows thus obtained BGF (unsharp-divided) images.

\cen{-- Fig. 1 --}

First of all, Fig. 1 reveals a bright bar-like feature near the nucleus.
The bar is elongated in the direction of PA=45\deg, and the major-axis length
is about $1'$ (=200 pc).
Fig. 2 shows an enlarged image of the bar in $R$ band.
The major-to-minor axial ratio is approximately 3.0,  much greater than that of
the central bulge component (axial ratio of 1.3 at PA=55\deg; Kinman 1965).
This fact implies that the bar feature is a distinct structure superposed on
the more round-shaped bulge component.

\cen{-- Fig. 2}

The apparently bar-like feature may be either due to a nearly edge-on view of a
small (rotating) stellar disk, or a real bar structure.
If it is a perfect edge-on disk, the thickness is approximately 60 pc for the
radius of about 100 pc (diameter $1'.0$).
However, if it lies in the same plane as the major M31 disk ($i=77\Deg$; tilt
angle of 13\deg from the line of sight), the thickness must be less than 10 pc.
In either rate, if the disk is tilted even slightly from the line of sight, the
disk must be unrealistically thin.
Therefore, we may reasonably assume that the feature is a real bar structure.

The $V$-band flux from this bar (after subtracting the bulge component) is
estimated to be $\sim 5$\% of the total flux within 100 pc ($30''$) radius of
the center.
The total magnitude of M31 bulge in 30$''$ of the nucleus is estimated to be
8.0 mag. in V band, and a galactic-extinction-corrected magnitude of 7.7 V mag.
This yields an absolute magnitude of -16.5 mag., and $3\times10^8$ solar
luminosity, for an assumed distance of 690 kpc.
For a possible mass-to-luminosity ratio of about 5 (in solar units) for the
bulge region, we obtain a total mass of $\sim 1.5\times10^9 \Msun$.
On the other hand, the dynamical mass of the bulge within 100 pc (30$''$) can
be estimated as
$M_{\rm dyn} \sim r(v^2+\sigma^2)/G \sim 7\times10^8\Msun$
for an ellipsoid with a velocity dispersion
$\sigma=170$ \kms and rotation $v= 40$ \kms (Kormendy 1988), where $G$ is the
gravitational constant.
This dynamical mass leads to a mass-to-luminosity ratio of 2 (in solar units).
We may thus estimate that the bar has a mass of $\sim2 - 4\times10^7\Msun$.

%mean V intensity within 30"=40 pixel of the center = 9.77 x 20 mag.star / pix
%total mag in 60"square = 9.77 x 6400 pix x 20mag star = 8.01 mag
%If Av=0.3, then mag=7.7
%Abs.mag = -16.5mag (690 kpc) (-4.83 (sun) = -21.3 mag times Lv(Sun))
%L=3.3E8 Lv(sun)
%M/L=10 sol.unit then, M=3.4E9 Msun

\v\noi{\it 2.3. Dark Lanes along the Bar}

This bar is associated with dark lanes on its leading edges with respect to the
rotation sense of the galaxy (anti-clockwise).
In order to show up the dark lanes, we applied another contrast-enhancing
technique, called the $\Theta$ relief method (Sofue 1993):
The original image was rotated  by 10\deg around the nucleus, and the rotated
image was subtracted from the original, which yielded a differentiation in the
azimuthal direction and enhances radial features embedded in the bright
background.
Fig. 3 shows the $\Theta$ relief image in the $B$ band for the central $2'.5
\times 2'.5$.
The dark lanes run in the direction at PA=60\deg, and are slightly curving in
the trailing sense with respect to M31's rotation.
The length of each dark lane is about $30''$ (100 pc).
In order to clarify if these dark features are due to interstellar extinction,
we obtained  color excess maps in Fig. 4.
The dark lanes are significantly redder than the surrounding regions with an
excess in the color index of $E(B-V) =0.035$ and $E(B-I)=0.065$.
The ratio of the two excesses, $E(B-V)/E(B-I) \sim 0.5$, confirms that the dark
lanes are due to interstellar absorption (Walker 1987).
The bright mini bar has about the same color, and therefore about the same
population, as the bulge stars.

\cen{-- Fig. 3, 4 --}

In order to estimate the `true' color excess of the dark lanes, which can be
related to interstellar extinction, we need to correct for the foreground
emission of the bulge.
For this we simply assume that the contribution of the foreground emission is a
half of the total bulge emission.
Then the dark lanes' color excess (with respect to the bulge light behind the
lanes) is estimated to be about twice the above obtained apparent excess
values:
$E'(B-V)\simeq 2E(B-V)=0.07;~ E'(B-I)\simeq 2E(B-I)=0.13$.
If we assume a ``normal'' conversion formula from the color excess to column
density of hydrogen gas, we can  estimate the column density of hydrogen atoms
to be  $4\times10^{20}$ atoms cm$^{-2}$
toward the lane, where we used the empirical relation obtained for interstellar
matter in the solar vicinity;
$N({\rm H+H_2})/E'(B-V)\sim 6\times 10^{21}$ atoms cm$^{-2}$ mag$^{-1}$ (Savage
and Mathis 1979).
{}From the column mass, we estimate an approximate mass of the dark lanes to be
only $\sim 10^4\Msun$.
However, as is discussed in section 3 in detail, this estimate does not
necessarily  give the right value in view of a much smaller column mass which
is  estimated from the dynamical mass of the clouds using  molecular line-width
information.

\v\noi {\it 2.4. Dark Clouds along ``Face-on Spirals"}

The most conspicuous in figure 3 are dark patches with high $E(B-V)$ in the
northern region of the nucleus, located at $\Delta \delta \sim 75'', \Delta
\alpha \sim -20''$,
where $\Delta \delta$ and $\Delta \alpha$ are offsets in declination and right
ascension with respect to the nucleus, respectively.
These ``red'' patches are clearly identified with dark patches in the
contrast-enhanced image in figure 1.
The size of individual clouds is typically 15$''$ (50 pc).
The high color excess indicates interstellar extinction by dust, and  we may
call these patches dark (dust) clouds.
The excess in the color index of the clouds over that of the bulge  reaches
about $ E(B-V)=0.10$ and $ E(B-I)= 0.22$ at the darkest (reddest) cloud.
This indicates again a normal interstellar extinction (Walker 1987).
For the same assumption as above, we estimated the clouds' color excess to be
$ E'(B-V)=0.20$ and $ E'(B-I)= 0.44$,
which corresponds to a hydrogen column density of about
$N({\rm H+H_2}) \sim 1.2\times 10^{21}$ atoms cm$^{-2}$.
For a size of 50 pc, its mass is estimated to be $\sim 3 \times 10^4\Msun$.
This value is consistent with that obtained by Gallagher and Hunter (1981).

Besides these conspicuous clouds, a number of dark clouds are distributed over
the $2'$-radius region around the nucleus, especially in the NE to N and in SW.
The distribution of these dark clouds  appears to trace spiral arms, which look
rather ``face-on''.
This spiral pattern appears independent of the major spiral structure of the
M31 disk that is closer to edge-on at an inclination angle of 77\deg.
We point out that this face-on spiral distribution shows a good coincidence
with the nearly face-on spiral pattern observed in the \halpha emission
(Ciardullo et al. 1988).
We here mention that the dark clouds recognized in our analysis are not
associated with any star forming regions, as it is shown from inspection of
the $B$-band images.

\sect{\bf 3. CO-Line Emission}

\v\noi{\it 3.1. CO Observations}

In order to investigate the mass and kinematics of the dark clouds found in the
CCD images, we performed $^{12}$CO ($J=1$-0) line observations has been
performed using the Nobeyama 45-m telescope.
The result and observational detail have been reported in Sofue and Yoshida
(1993).
We here briefly summarize the result.

\v\noi{\it 3.3. Upper Values for CO Emission along the Mini Bar}

We observed the dark lanes associated with the mini bar at  $X=-30, -15,~ 0,~
15, 30''$ and $Y=0''$ as well as at $X=0'', Y=15''$, where $X$ and $Y$ are
coordinates taken along a line crossing the nucleus at PA=67\deg ($X$ positive
toward NE, and $Y$ positive toward NW).
The obtained spectra for the individual points and their sum are reproduced
from Sofue and Yoshida (1993) in Fig. 5.
No significant emission was detected, yielding an upper limit of about 20 mK
\tmb.
For an assumed velocity width of about 100 \kms, this gives an upper limit to
the molecular hydrogen mass of  $\sim 1\times10^5\Msun$ within the observed
area (mini bar).
The weaker CO emission compared to the northern dark clouds as described below
is consistent with the smaller $E(B-V)$ and $E(B-I)$ values near the mini bar,
which indicates less extinction and therefore smaller amount of molecular gas.
We mention, however, that we cannot deny the possibility of non-detection due
to a possibly broad line profile, which often occurs near the nucleus due to
large velocity dispersion and high rotation velocity.

\cen{-- Fig. 5 --}

\v\noi{\it 3.2. CO Emission from Northern Dark Clouds}

We observed several positions around the northern dark clouds as shown in Table
1.
Figure 6 shows the obtained CO line spectra for the six positions as well as a
composite spectrum obtained by integrating the six spectra.
The CO emission has been detected at all positions at \tmb = 100 to 180 mK
except for D .
The mean velocities of the emission are about $-220$ \kms at around point C,
and about $-250$ \kms at F.
These velocities agree with the velocities of $\sim -200$ to $-250$ \kms as
derived from optical spectroscopy of the [OII] and [NeIII] lines at $\sim 80''$
from the nucleus at PA=$-7\Deg$ and $-38\Deg$ (to NE), respectively (Ciardullo
et al. 1988).
The observed data are summarized in Table 1.

\cen{-- Fig. 6 --}

The darkest cloud in the northern complex  has an optical size of $10''$ in
diameter, or the radius is  $r \sim 17$ pc.
The CO velocity width of $\sigma_v \sim 30$ \kms has been observed, which
yields a  virial mass of the cloud of
$M_{\rm vir} \sim 9 \times 10^5 \Msun$,
where $G$ is the gravitational constant.
The total mass involved in the dark complex is then estimated to be some
$10^6\Msun$.
Thus, the dark cloud complex has a mass comparable to a giant molecular cloud.
We stress that the dark complex is most conspicuous, and therefore probably
most massive, within the central few hundred pc of M31's nucleus.
This shows a striking contrast to the nuclear disk of our Galaxy, where giant
molecular complexes by one or two orders of magnitude more massive have been
observed.

The column density of \htwo of this cloud is then estimated to be
$N_{\rm H_2} 6 \times 10^{22}$ \htwo cm$^{-2}$.
After applying a correction for the beam-dilution effect
we obtain the CO intensity toward the dark cloud  to be
$ I_{\rm CO} \sim 20 $ K \kms.
{}From these, we derive a conversion factor from CO intensity to \htwo column
density as
$ X=N_{\rm H_2}/I_{\rm CO} \sim 3 \times 10^{21} $ \htwo cm$^{-2}$/K \kms (= 50
\msun pc$^{-2}$/K \kms).
This value is by a factor of ten greater than that  for molecular clouds in our
Galaxy as estimated from a similar virial-mass method:
$ \sim 3.6 \times 10^{20}$ \htwo cm$^{-2}$/K \kms (Sanders et al. 1984).
The is anomalous conversion factor is similar to that observed in the Small
Magellanic Clouds (Rubio 1991), and also the weak CO emission is similar to
that observed molecular clouds in the Large Magellanic Cloud (Booth and de
Glaauw 1991).

We also find that the gas-color-excess ratio is anomalous:
We obtain a gas-to-color excess ratio of
$  N_{\rm H_2} / E ( B - V ) \simeq  6 \times 10^{23}$ atoms cm$^{-2}{\rm
mag.}^{-1}$.
This is almost by two orders of magnitude greater than that for inter-cloud
value in the solar vicinity, $\sim 6 \times 10^{21}$ atoms cm$^{-2}{\rm
mag.}^{-1}$, which might increase significantly for dust in dense clouds
(Savage and Mathis 1979).

At any rate,  the observations indicate a anomalously small amount of molecular
gas in M31's bulge compared to those observed in the central few hundred pc in
other Sb galaxies like the Milky Way (Dame et al 1990) and NGC 891
(Garcia-Brillo et al. 1992; Sofue and Nakai 1993).
In view of such a small amount of molecular gas, it is not surprising that no
indication of star formation can been recognized from an inspection to the $B$-
and $V$-band images (Fig. 1).

The small CO intensity-to-gas mass ratio indicates that the heating source of
the clouds is lacking, and, hence, the star formation rate in the central few
hundred pc is very low.
The anomalously small color excess indicates that the dust content per unit
mass of the gas is small, and hence gas is dust deficient.
Therefore, we may conclude that the molecular clouds in the central region is
more ``primeval'' and appear to be new comers rather than a long-living
polluted gas.
{}From this fact, we suggest that the gas clouds could be originated by infall
from external galaxies of Magellanic type, which we discuss in the next section
in some detail.

\sect{\bf 4. Discussion}\v

The N to NE and SW dark clouds in M31's center appear to trace a ``face-on"
spiral in a positional as well as morphological coincidence with those of
ionized gas (Ciardullo et al. 1988).
These spiral features appear to be connected to the more central bar-like
structure, which is associated with dark lanes on its leading edges.
All these suggest a face-on barred spiral structure composed of dark clouds as
well as the central mini stellar bar.

The origin of such a  barred spiral structure  of ionized as well as molecular
gases in the central few hundred pc, which is highly-tilted with respect to the
M31 main disk, is a mystery, providing an interesting subject.
A possible scenario involves ejection of ionized gas from M31's bulge via a
galactic wind which was driven by energy pumped into the interstellar matter by
supernova explosions (Ciardullo et al. 1988).
However, since the central 1 kpc region contains very little gas and no nuclear
gaseous disk is present (Sofue and Yoshida 1993), such an explosion-driven wind
must blow spherically, and, hence, it seems difficult to collimate the wind in
the out-of-plane spiral.

Recently we proposed a ram-pressure stripping-and-accretion model of  gaseous
debris from the companions, M32 and NGC 205 (Sofue 1994).
A merger galaxy, which is suggested from the existence of double nuclei (Lauer
et al 1993), could be an alternative origin of the gaseous debris.
In this scenario we assume that M32 and NGC 205 or the merger galaxy were
companions which had contained much interstellar gas in the past.
The model shows that the HI and molecular gas clouds are both stripped from the
companions, and are captured by M31's gravity.
The captured clouds are then rapidly accreted toward the galaxy disk along
highly-tilted (polar) orbits within $\sim 10^9$ yr.
The clouds, then, make a polar rotating disk in the central region, in which
the clouds are further accreted toward the nucleus along spiral orbits in $\sim
10^8$ yr.

Some of the clouds would remain in the form of giant molecular clouds, as they
were observed here in CO and as dark clouds.
If the companion galaxies, including the possible merger galaxy, were of
``primeval'' like galaxies similar to the Magellanic Clouds, the anomalously
small ratios of the CO intensity and color excess to cloud mass can be
explained.
During the accretion, the clouds will be partially heated and ionized due to
friction with the ambient interstellar gas, which would produce the \halpha
spiral features.
However, the rest of the clouds remained low temperature, since no  activity is
present (section 2), which is required to heat up the gas to radiate strong CO
emission.
The accretion of massive gas clouds,  and possibly the merger of a galaxy
(Lauer et al 1993), would have caused a tidal disturbance in the bulge, and
disrupted the nuclear gas disk of M31.
At the same time, the gravitational disturbance would have caused the central
mini bar structure.
Furthermore, a bar-induced accretion of interstellar gas might be somehow
related to the formation of a massive black hole suggested to be present at the
nucleus (Dressler and Richstone 1988; Kormendy 1988; Lauer et al 1993).

\sect{\bf References}

\re Booth, R. S., and de Graauw, Th.  1991, in {\it The Magellanic Clouds}, ed.
R. Haynes, and D. Milne (\kluwer), p.415.

\re Brinks, E. and Bajaja, E. 1986, \aa, 169, 14.

\re Ciardullo, R., Rubin, V. C.,Jacoy, G. H., Ford, H. C., Ford, Jr., W. K.
1988, \aj, 95, 438.

\re Combes, F., Encrenaz, P., Lucas, R., and Weliachew, L. 1977, \aa, 61, L7.

\re {Dame, T. M., Ungerechts, H., Cohen, R. S., de Geus, E. J., Grenier, I. A.,
May, J., Murphy, D. C., Nyman, L. -A, and Thaddeus, P.  1987, \apj, 322, 706. }

\re Dressler, A., and Richstone, D. O. 1988 \apj, 324, 701.

\re Gallagher, J. S, and Hunter, D. A. 1981 \aj, 86, 1312.

\re Garcia-Burillo, S., Gu{\'e}lin, M., Cerhicharo, J., Dahlem, M. 1992, \aa,
{\bf 266}, 210.

\re Habing, H. J., Miley, G., Young, E., Baud, N, Bogges, N., Clegg, P. E., de
Jong, T., Harris, S., Raimond,E., Rowan-Robinson, M., and Soifer, B. T.  1984,
\apj, 278, L59.

\re Hodge, P. W. 1980 \aj, 85, 376.

\re Johnson, H. M., and  Hanna, M. 1972, \aj, 174, L71.

\re Kinman, T. D. 1965, \apj, 142, 1376.

\re Koper, E., Dame, T. M., Israel, F. P., Thaddeus, P. 1991, \apjl, 383, L11.

\re Kormendy, J. 1988, \apj, 325, 128. % (M31 BH)

\re Light, E. S., Danielson, R. E., Schwarzschild, M. 1974,\apj, 194, 257.

\re McEroy, D. B. 1983, \aj, 270, 485.

\re Roberts, M. S., and Whitehurst,  R. N., 1975, \aj, 26, 483.

\re Rubio, M. 1991, in {\it The Magellanic Clouds}, ed. R. Haynes, and D. Milne
(\kluwer), p.429.

\re Savage, B. D. and Mathis, J. S. 1979, \araa, 17, 73.

\re Sofue, Y. 1993, \pasp, 105, 308.

\re Sofue, Y. 1994, \apj, March issue, in press.

\re Sofue, Y., and Nakai, N. 1993, \pa, 46, 139.

\re Sofue,Y., and Reich, W. 1979, \aas, 38, 251.

\re Sofue, Y., and Yoshida, S. 1994, \apjl, in press.

\re \so and \wa\ 1993, \aa, 273, 79.

\re Stark, A. A.  1985, in {\it The Milky Way Galaxy}, IAU Symp. No. 106, ed.
H. van Woerden, R. J. Allen, and W. B. Burton (Reidel Pub. Co., Dordrecht), p.
445.

\re Sanders, D. B., Solomon, P. M., and Scoville, N. Z. 1984, \apj, 276, 182.

\re Walker, G. 1987, in {\it Astronomical Observations} (Cambridge University
Press), p. 17.

\endpage
\noi Table 1. Positions of CO-line detctions$^\dagger$.
\v
\settabs 5\columns
\hrule \vskip 0.5mm \hrule
\v
\+ Position & RA-offset & Dec-offset & Peak \tmb(mK) & \ico (K \kms) \cr
\v
\hrule
\v
\+ A \dotfill & $-23''$ & $90''$ & 120$\pm20$ & 5.4$\pm 1.3$ \cr
\+ B \dotfill & $-8''$ & $75''$ & 120 & 5.5 \cr
\+ C \dotfill & $-23''$ & $75''$ & 180 & 9.0 \cr
\+ D \dotfill & $-38''$ & $75''$ & $<40$ & ... \cr
\+ E \dotfill & $-23''$ & $60''$ & 100 & 4 \cr
\+ F \dotfill & $-8''$ & $45''$ & 160 & 6.4 \cr
\v
\hrule
\v
$\dagger$ RA and Dec offsets  referred to the M31 nucleus at
RA = 00h 40m 00.1s, Dec = $40\Deg59'42'.7$ (1950)(NED 1992).
The errors in \tmb and \ico are $\pm20$ mK and $\pm 1.3$ K \kms, respectively.

\endpage

\noi Figure Captions

\v\v
\r Fig. 1: Background-filtered $B$ (left; stars are subtracted) and $I$-band
(right panel) images of the central $4'.3\times4'.3$ region of M31 taken with a
CCD equipped on the Kiso 105-cm Schmidt telescope.
The original images were divided by Gaussian smoothed images with HPBW of 10
pixels ($7''.5$) (``unsharp-divided'' images).
The darkest region is 0.18 mag darker than the average in $B$ band, and 0.07
mag in $I$ band, which, however, do not represent true darkening, because
smoothed cloud structures have been subtracted as well.
For a quantitative discussion of extinction, see color excess maps in Fig. 4.
 (N to the top; E to the left.)

\v
\r Fig. 2: BGF (unsharp-divided) image of the  mini bar in $R$ band.
The central $1'.25$ squared region is shown.
Contours are drawn every $-0.021$ mag. starting at $-0.01$ mag. relative to the
smooth background. %1.0 to 1.78 interval 0.02

\v
\r Fig. 3: $\Theta$-relief $B$ band image of dark lanes associated with the
central ``mini bar" near the nucleus of M31.
The central $1'.25$ squared region (the same as Fig. 2) is shown.
The contours are drawn every 0.01 mag. starting from 0.01 mag.  relative to the
background (darker toward contour peaks), but for a more quantitative
information can be obtained by
color excess maps in Fig. 4.
%b1r20, 0.01 int
\v
\r Fig. 4: Color excess maps for $E(B-V)$ (left panel) and $E(B-I)$ (right
panel). The central $3'.76$ squared region is shown.
The contours are drawn at interval of 0.016 mag., starting at 0.016 mag., in
excess over the background color index of the bulge,
The reddest cloud has an index excess of $E(B-V)=0.10$ and $E(B-I)=0.22$.

\v
\r Fig. 5: CO-line spectra toward the central mini-bar and its dark lanes
(upper panel) obtained with the Nobeyama 45-m telescope, and a composite
spectrum (lower)  by integrating the spectra in the upper panel.
Indicated positions are $(X,Y)$ offsets in arcsec relative to the nucleus,
where $X$ is the axis along a line at PA=67\deg ($X$ positive to NE; $Y$
positive to NW).
No significant emission was detected.

\v
\r Fig. 6: CO-line spectra obtained for the northern dark clouds (upper panel),
and their composite spectrum (lower).

\bye